\documentclass[aps,pra,showpacs%
        ,amsmath%
        ,superscriptaddress%
        ,reprint%
        ,floatfix%		Comment out for submission
        ]{revtex4-1}
\usepackage[pdftex]{graphicx} % arXiv

\usepackage{amssymb}
\usepackage{amsmath} % incompatible with iopart.cls

\usepackage[mathscr]{eucal}

\newcommand{\mL}{m_{\rm L}}
\newcommand{\mR}{m_{\rm R}}
\newcommand{\nL}{n_{\rm L}}
\newcommand{\nR}{n_{\rm R}}

\newcommand{\bra}[1]{\langle{}#1{}|}
\newcommand{\ket}[1]{|{}#1{}\rangle}
\newcommand{\bracket}[2]{\langle{}#1{}|{}#2{}\rangle}

\newcommand{\CX}{C_{\rm X}}
\newcommand{\CY}{C_{\rm Y}}

\begin{document}

%%% Comment out below for submission: put revision No.
% \preprint{{\sf 
%     DRAFT
%     %: NOT FOR DISTRIBUTION
%     % CLOSING AT1
%     %{\AT1 ver.AT1
%     (2015-10-23)
%     %}
%   }}
%\date{} % Comment out for submission

%%%
%%% Title page
%%%
%\title{Adiabatic excitation of a confined one-dimensional particle with a variable $\delta$-wall}
\title{Adiabatic excitation of a confined particle in one dimension with a variable infinitely sharp wall}

%\author{Sho Kasumie and Atushi Tanaka}
% \Address{$^1$ %NOPREPRINT%
%   Department of Physics, Tokyo Metropolitan University,
%   Hachioji, Tokyo 192-0397, Japan}
% \ead{tanaka-atushi@tmu.ac.jp}%NOPREPRINT%
% \address{$^2$ %NOPREPRINT%
%   Laboratory of Physics, Kochi University of Technology, Tosa Yamada, 
%   Kochi 782-8502, Japan} 
% \ead{taksu.cheon@kochi-tech.ac.jp}%NOPREPRINT%

\author{Sho Kasumie}
\altaffiliation[Present address: ]{%
Okinawa Institute of Science and Technology, Okinawa 904-0495, Japan 
}%\altaffiliation
\affiliation{Department of Physics, Tokyo Metropolitan University,
  Hachioji, Tokyo 192-0397, Japan}
\author{Manabu Miyamoto}
\affiliation{Department of Physics, Waseda University, 3-4-1 Okubo, Shinjuku-ku,Tokyo 169-8555, Japan}
\author{Atushi Tanaka}
%\email[]{tanaka-atushi@tmu.ac.jp}
\homepage[]{\tt http://researchmap.jp/tanaka-atushi/}
\affiliation{Department of Physics, Tokyo Metropolitan University,
  Hachioji, Tokyo 192-0397, Japan}

\date{\today}

\begin{abstract}
It is shown that adiabatic cycles excite a quantum particle, which is confined in a one-dimensional region and is initially in an eigenstate. During the cycle, an infinitely sharp wall is applied and varied its strength and position. After the completion of the cycle, the state of the particle arrives another eigenstate. It is also shown that we may vary the final adiabatic state by choosing the parameters of the cycle. With a combination of these adiabatic cycles, we can connect an arbitrary pair of eigenstates. Hence, these cycles may be regarded as basis of the adiabatic excitations. A detailed argument is provided for the case that  the particle is confined by an infinite square well. This is an example of exotic quantum holonomy in Hamiltonian systems.
\end{abstract}

\pacs{03.65.-w,03.65.Vf}
%% PACS2010
%% 03.65.-w 	Quantum mechanics
%% 03.65.Vf     Phases: geometric; dynamic or topological 

\maketitle

\section{Introduction}
\label{sec:Introduction}

An adiabatic process, also referred to as an adiabatic passage, 
offers a simple and robust way to control a quantum
system~\cite{AP}. An adiabatic passage connects a stationary state of the
initial system with another stationary state of the final system 
%{\AT1%
through a slow variation of an
%by slowly varying the
%}%\AT1%
external field,
%{\AT1\sout{%
%of the system, 
%}}%{\AT1\sout{%
for example. This has been
investigated both experimentally and theoretically to a variety of
microscopic systems, e.g., atoms and 
molecules~\cite{Rice2000,Shapiro2012,Shore2011}.

We here focus on the adiabatic passage along a closed cycle in the
adiabatic parameter space. This corresponds to the case, for example,
where the external field is applied only during a cycle, and is off
before and after the cycle.
One may expect that the adiabatic cycle induces no change, since
there is no external field to keep the final state away from the
initial one.
This argument, however, has counterexamples. A famous one is
the appearance of the geometric phase factor~\cite{Berry-PRSLA-392-45}, 
which is also referred to as the quantum holonomy~\cite{Simon-PRL-51-2167}. 
Furthermore, an adiabatic cycle may deliver a stationary state into another
stationary state. This can be regarded as an permutation of eigenspaces.
Since this is analogous with the quantum holonomy,
we call such an excitation due to adiabatic cycles
an exotic quantum holonomy (EQH).
% Make simple the following
Examples of EQH 
are reported through theoretical 
studies~\cite{Cheon-PLA-248-285,Tanaka-PLA-379-1693}.

In atomic and molecular systems, there are many studies of 
population transfers with the oscillating classical electromagnetic field
both in theory and experiments. Among them, 
stimulated Raman adiabatic passage 
(STIRAP)~\cite{Rice2000,Shapiro2012,Shore2011}
employs an adiabatic cycle made of quasienergies, which is the counterpart
of eigenenergy in periodically driven systems~\cite{Zeldovich-JETP-24-1006}. 
Its adiabatic cycle
passes through a crossing point of 
quasienergies to realize the adiabatic
excitation (or its inverse)~\cite{Guerin-PRA-63-031403,Yatsenko-PRA-65-043407}.
Hence, STIRAP can be considered as an example of EQH.
On the other hand, it has been shown that,
even in the absence of level crossings,
periodically driven systems may exhibit
EQH~\cite{Tanaka-PRL-98-160407,Miyamoto-PRA-76-042115}.

In this manuscript, we offer another example of EQH. This example consists
of a particle confined in one-dimensional region, where a slowly varying 
%{\AT3%
wall
%potential 
%}%\AT3%
is applied
%{\AT1%
to make cycles.
%{\AT3%
We assume that the time-dependent wall is described by a potential, which
is propotional to the $\delta$-function.
%}%\AT3%
% {\AT2%
% We choose the shape of time-dependent part of the potential as
% infinitely sharp one, which 
% }%{\AT2%
% %The shape of time-dependent part of the potential 
% %}%\AT1%
% % {\AT1\sout{%
% % An additional potential whose shape 
% % }}%\AT1\sout{%
% is proportional to the $\delta$-function.
%{\AT1%
We call it a $\delta$-wall.
%}%\AT1%
%{\AT1\sout{%
%is applied 
%}}%\AT1\sout{%
%

Since our model is described by a slowly time-dependent Hamiltonian,
EQH is governed by the parametric dependence of eigenenergies, instead
of quasienergies. This is in contrast with the examples with
oscillating external field mentioned above. Also, 
%{\AT1\sout{%
%the $\delta$-wall
%in 
%}}%\AT1\sout{%
our model is far simpler than the known examples of EQH 
%{\AT1%
in Hamiltonian systems, 
%}%\AT1%
for example, a particle under
the generalized pointlike
potential~\cite{Cheon-PLA-248-285} and a quantum graph with a
$4$-vertex~\cite{Cheon-ActaPolytechnica-53-410}.
%
%{\AT1%
Experimental 
%Hence, experimental 
%}%\AT1%
realizations of our model may be feasible within 
the current state of art, as is suggested by the realization of optical 
box trap made of two thin walls~\cite{Meyrath-PRA-71-041604},

%{\AT2%
We show that the adiabatic cycles of the present model may have 
qualitatively distinct result.
%We show that the present model has topologically distinct nontrivial 
%adiabatic cycles.
Namely, by adjusting the parameters of the adiabatic cycle (i.e., the initial and
final positions of the $\delta$-wall), we may vary 
the final stationary state. Also, by combining these cycles, we can
connect an arbitrary adiabatic state. 
In this sense, these cycles form the basis of adiabatic excitations.
In particular, we
%We 
%}%\AT2%
%
will examine two 
%{\AT2%
kinds of
%}%\AT2%
adiabatic cycles, which induces
different permutations of eigenspaces.
One is denoted as $\CX$,
which involves an insertion, a move, and a removal of the
$\delta$-wall.
Although this resembles a simple thermodynamic
process~\cite{Kim-PRL-106-070401}, 
its consequence would have no similarity in thermodynamics.
Namely, we will show that the final state of the 
adiabatic cycle $\CX$ is different from the initial one, depending on 
the positions of the wall.
In other words, $\CX$ may induce an excitation
of the system.
Another example $\CY$
involves an insertion, a flip, and a removal 
of the wall.
The eigenspace permutation induced by $\CY$ resembles the one found 
in the Lieb-Liniger model~\cite{Lieb-PR-130-15}, where all 
eigenspaces are excited at a time~\cite{Yonezawa-PRA-87-062113}.

The plan of this manuscript is the following. We introduce our model
in Section~\ref{sec:model} and 
%{\AT2
provide an exact analysis of
%examine 
%}%\AT2
the application of 
the $\delta$-wall in Section~\ref{sec:apply}. The adiabatic
cycles $\CX$ and $\CY$ are examined in Sections~\ref{sec:CX} and 
\ref{sec:CY}, respectively. Also in Sections~\ref{sec:CX}, we
show that a combination of $\CX$'s 
adiabatically connects an arbitrary pair of eigenstates. 
Section~\ref{sec:conclusion} is dedicated to conclusion.

\section{A confined 1-d particle with a $\delta$-wall}
\label{sec:model}
In this section, we introduce 
a confined 1-d particle with a $\delta$-wall.
Suppose that a particle is confined within a one-dimensional region,
whose position is denoted by $x$.
The particle is described by a Hamiltonian
$H_0=T+V_0(x)$, where $T\equiv p^2/2$ and $V_0(x)$ are the kinetic term 
and the confinement potential, respectively, and the particle mass is 
chosen to be unity.

Although the following argument is applicable to a wide class of 
confinement potentials $V_0(x)$ as long as $H_0$ has several bound states
and has no spectral degeneracy,
it is convenient 
to employ an infinite square well to explain the concept.
We assume $V_0(x)=0$ for $0\le x \le L$, and $V_0(x)=\infty$ otherwise.
As for the infinite square well,
the eigenenergies of $H_0$ are
$E^{(0)}_n = (\hbar \pi n /L)^2 /2$
($n=1,2,\dots$).
Let $\ket{n}$ denote the corresponding normalized stationary state.

We introduce an additional $\delta$-wall, where 
%Our aim is to connect the stationary states of $H_0$
%by adiabatic cycles, where an additional $\delta$-wall is introduced.
the whole system is described by the Hamiltonian
\begin{equation}
  \label{eq:def_H}
  H(g, X)\equiv H_0 + g\delta(x-X)
  .
\end{equation}
We assume that we may vary 
the strength $g$ and position $X$ of the wall.
Let $E_n(g,X)$ be an eigenenergy of $H(g,X)$
for $-\infty < g < \infty$, i.e., 
\begin{equation}
  \label{eq:eigen}
  H(g,X)\ket{n(g,X)}=E_n(g,X)\ket{n(g,X)}
  ,
\end{equation}
where we impose $E_n(0,X)=E_n^{(0)}$ and $\ket{n(0,X)}=\ket{n}$.
We refer that an infinite square well with a $\delta$-wall was
examined in Refs.~\cite{Flugge1971,Ushveridze-JPA-21-955}.

\section{Adiabatic application of $\delta$-wall}
\label{sec:apply}
%\paragraph{Adiabatic Insertion of the wall}
We examine the adiabatic insertion or removal of the $\delta$-wall,
where the strength $g$ is varied while the position $X$ is fixed.
%We will show that an appropriate choice of $X$ excludes the crossing
%of eigenenergies that are smaller than a certain cutoff, for an 
%arbitrary $g$. 
The following analysis tells us how the eigenstates and eigenvalues
of the unperturbed system (i.e, $g=0$) are connected to
the ones in the limit $g\to\infty$.

First, we examine the ``exceptional'' case.
Suppose that the node of the eigenfunction $\bracket{x}{n_0(g_0,X)}$
coincides with the position of the $\delta$-wall, i.e.,
\begin{equation}
  \label{eq:n0node}
  \bracket{X}{n_0(g_0,X)}=0
  , 
\end{equation}
which implies 
$\ket{n_0(g,X)} = \ket{n_0}$ and $E_{n_0}(g,X)= E_n^{(0)}$
for an arbitrary $g$, as is shown in the following.
Since $H(g,X)$ satisfies 
\begin{equation}
  \label{eq:Hsplit}
  H(g,X) = H(g_0,X) + (g - g_0)\delta(x-X)
\end{equation}
from Eq.~(\ref{eq:def_H}),
we obtain
$
H(g,X)\ket{n_0(g_0,X)}=
E_{n_0}(g_0,X)\ket{n_0(g_0,X)}
+ (g - g_0)\delta(x-X)\ket{n_0(g_0,X)}
$.
For an arbitrary normalized state $\ket{\phi}$, we find
\begin{align}
  \bra{\phi}H(g,X)\ket{n_0(g_0,X)}
  = E_{n_0}(g_0,X)\bracket{\phi}{n_0(g_0,X)}
  ,
\end{align}
because of Eq.~\eqref{eq:n0node}.
Hence we conclude 
$H(g,X)\ket{n_0(g_0,X)}=E_{n_0}(g_0,X)\ket{n_0(g_0,X)}$,
which implies $\ket{n_0(g,X)} = \ket{n_0}$ and $E_{n_0}(g,X)= E_n^{(0)}$.
We note that the condition Eq.~\eqref{eq:n0node} is equivalent
to $\bracket{X}{n_0}=0$.

Since the analysis of exceptional levels are completed, we exclude them
in the following. Namely, we focus on the levels whose 
unperturbed eigenstates satisfy $\bracket{X}{n}\ne0$, which 
ensures that $\bracket{X}{n(g,X)}\ne0$ holds for an arbitrary $g$.

We show that there is no level crossing among these levels.
Let us examine a pair of the levels $n$ and $n'$, 
where $E_n^{(0)} < E_{n'}^{(0)}$ is assumed.
We find from Eqs.~(\ref{eq:eigen}) and~(\ref{eq:Hsplit}),
\begin{eqnarray}
  \label{eq:nonlocal}
  &
  \left\{E_{n}(g,X)- E_{n'}(g',X)\right\}
  \bracket{n(g,X)}{n'(g',X)}  
  \nonumber\\ 
  &{}\qquad\qquad
  = 
  (g-g')\bracket{n(g,X)}{X}\bracket{X}{n'(g',X)}
  .
\end{eqnarray}
When $g\ne g'$ holds, 
we obtain
\begin{equation}
  \left\{E_{n}(g,X)- E_{n'}(g',X)\right\}
  \bracket{n(g,X)}{n'(g',X)}  
  \ne 0
  ,
\end{equation}
because of $\bracket{n(g,X)}{X}\bracket{X}{n'(g',X)}\ne 0$.
%% AT: added an explanation
Namely, we find 
\begin{equation}
  \label{eq:E_LongRange}
  E_{n}(g,X)\ne E_{n'}(g',X) 
  \quad\text{for $g\ne g'$}
  .  
\end{equation}
Since $E_{n}(g,X)$ monotonically increases with $g$ 
%{\AT1%
strictly, 
%}%\AT1%
we conclude
\begin{equation}
  \label{eq:Erange}
  E_{n}(g,X) < E_{n}(\infty,X) 
  \le E_{n'}(-\infty,X)  < E_{n'}(g,X)
\end{equation}
for an arbitrary $-\infty < g < \infty$.
Thus 
% {\AT1\sout{%
% we conclude that 
% }}%\AT1\sout{%
there is no level crossing between $n$-th and $n'$-th
levels.

A simple condition that ensures the absence of level crossing
among the levels $n < n^*$, where $n^*$ defines a cutoff, is
$\bracket{X}{n}\ne0$ holds for all $n < n^*$.
For the infinite square well, all unperturbed eigenstates satisfy 
this condition as long as $X/L$ is an irrational number.

\section{Cycle with expansion/compression}
\label{sec:CX}
\begin{figure}[bt]
  \centering
  \includegraphics[%
	width=8.7cm% for double columns
        ]{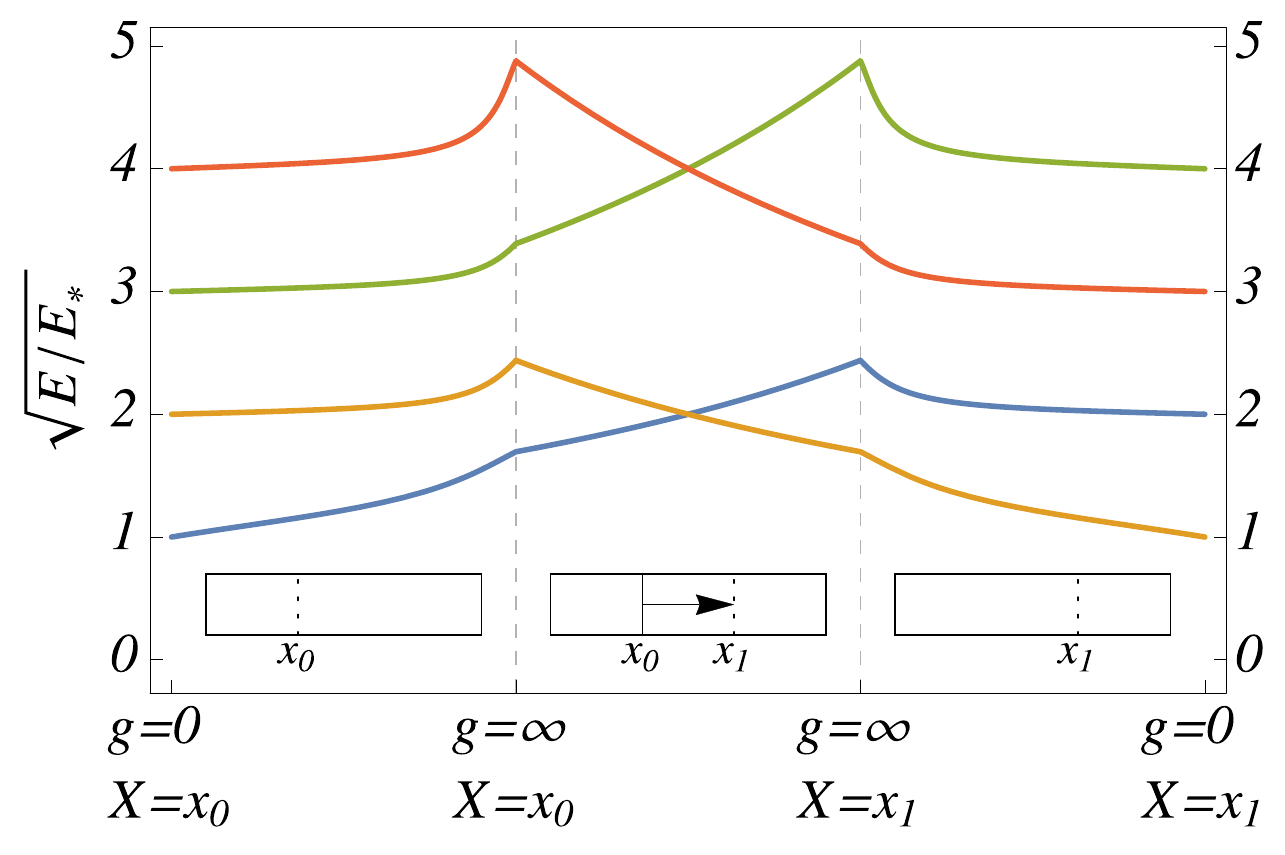}
  \caption{
    (Color online)
    Parametric evolution
    of the eigenenergies of a particle confined in an infinite square well, 
    along the closed cycle $\CX(x_0, x_1)$.
    The lowest four eigenenergies are shown.
    We choose $x_0 = 0.41 L$ and $x_1= 0.59 L$, where $L$ is the size of 
    the well. $E_*\equiv\frac{1}{2}(\hbar \pi/L)^2$ is the ground energy
    of the unperturbed system.
    The insets illustrate the position of the $\delta$-wall in each process
    schematically.
    (Left) A $\delta$-wall is initially placed at $X=x_0$ (dotted line)
    with its strength $g=0$.
    While increasing $g$ to $\infty$, there is no level crossing.
    (Middle) The impermeable $\delta$-wall (i.e., $g=\infty$) is moved from
    $x_0$ (full line) to $x_1$ (dashed line).
    A level crossing occurs at $X=L/2$.
    (Right) While $X=x_1$ is kept fixed, $g$ is decreased to
    $0$ to finish the cycle. There is no level crossing in this process.
    As a result, the cycle interchanges the ground and first-excited energies.
    The second and third-excited eigenenergies are also interchanged.
    The adiabatic time evolution follows along these lines.
    %{\AT1%
      We note that the horizontal axis of the left and right part is 
      linear in $\tan^{-1}g$.
    %}%\AT1%
  }
  \label{fig:n=1}
\end{figure}

We examine an adiabatic cycle $\CX(x_0, x_1)$, which 
consists of an insertion, a move, and a removal of the
$\delta$-wall. In particular, we impose that, during the second process,  
the $\delta$-wall is impermeable, i.e., $g=\infty$, to completely divide
the confinement well into two regions. 

The key to realize the adiabatic excitation through $\CX$ 
is to utilize the level crossing during the second process.
The same concept has been utilized 
in Refs.~\cite{Guerin-PRA-63-031403,Yatsenko-PRA-65-043407,Cheon-PLA-374-144} 
to realize the adiabatic excitations along cycles.
Although the level crossings are generally fragile against perturbations
according to  Wigner and von Neumann's theorem~\cite{WvN},
this theorem is inapplicable to our case since the system is completely
divided into two parts.
In reality, if we take into account the imperfection of the impermeable wall,
e.g. the effect of tunneling, 
the spectral degeneracy may be lifted.
%the level crossing may be lifted.
There we need to resort to the diabatic evolution to go across 
the avoided crossing in order to approximately realize the adiabatic
excitation~\cite{Cheon-PLA-374-144}.

We assume that the system is initially 
in a stationary state $\ket{n}$, i.e., the $n$-th excited state 
of the initial system $\hat{H}_0$.
We will show that the final state is
the $(n+1)$-th excited state,
if we appropriately choose $x_0$ and $x_1$,
i.e., $\CX(x_0, x_1)$ delivers the initial stationary
state to its higher neighboring state.
For simplicity, 
we assume in this section that the confinement potential is the 
infinite square well, and $x_j/L$ ($j=0,1$) is irrational.

A precise definition of  $\CX(x_0, x_1)$ is shown.
First, the wall is adiabatically inserted at $x_0$. Namely, the
strength $g$ is adiabatically increased from $0$ to $\infty$, while
the position is kept fixed at $X=x_0$. 
%We will explain below
%the condition that the quantum number $n$ is kept fixed during this 
%process. 
After the completion of this
process, the $\delta$-wall is impermeable, i.e., the confinement is
divided into two regions. 
Second, the impermeable wall is moved adiabatically from $x_0$ to
$x_1$. If $x_0 < x_1$ ($x_0 > x_1$) holds, we say that the left
(right) well is expanded while the other well is compressed.
Third, the wall is adiabatically removed, where the strength $g$ is
adiabatically decreased from $\infty$ to $0$ while the position is
kept fixed at $X=x_1$. After the completion of this process, the
system is described by the unperturbed Hamiltonian $H_0$ again. 

During the first process, the 
%{\AT1%
state vector of the
%}%\AT1%
system is 
% {\AT1\sout{%
% in 
% }}%\AT1\sout{%
$\ket{n(g, x_0)}$,
%{\AT1%
up to a phase factor.
%}%\AT1%
There is no level crossing $0\le g < \infty$, since we choose 
$x_0/L$ is irrational (see Sec.~\ref{sec:apply}).
Hence the system is always in $n$-th excited state. This implies that
the system is also in the $n$-th state right before the second process.
Similarly, the system is in 
the $(n+1)$-th excited state right after the second process.

We explain how the $n$-th and $(n+1)$-th states are connected
by the second process,
i.e., the moving the impermeable wall from $x_0$ to $x_1$,
%{\AT1%
through
%}%\AT1%
% {\AT1\sout{%
% We utilize 
% }}%\AT1\sout{%
a level crossing.
We examine this process 
%{\AT1%
by introducing 
%}%\AT1%
%in terms of
another quantum numbers $\mL$ and $\mR$ 
% {\AT1\sout{%
% defined below for 
% }}%\AT1%
%{\AT1%
of
%}%\AT1%
the separated
%{\AT1%
systems.
%}%\AT1%
% {\AT1\sout{%
% system, instead of $n$. 
% Namely, we solve the condition 
% for the level crossing of energy levels corresponding to $\mL$ and $\mR$.
% }}%\AT1\sout{%

Since we assume that $V_0(x)$ is the infinite square well,
the system is divided into two 
infinite square wells during the second process.
The left and right wells are 
placed at $0 \le x \le X$ and $X \le x \le L$, respectively.
We introduce two quantum numbers $\mL{}$ and $\mR{}$, which
describes the particle confined in 
the left and right wells, respectively, under the presence of 
the impermeable wall.
The eigenenergies are
$E_{{\rm L},\mL{}}(X) = (\hbar \pi \mL{} /X)^2 /2$
and $E_{{\rm R},\mR{}}(X) = (\hbar \pi \mR{} /(L-X))^2 /2$.

We examine the level crossing consists of $\mL{}$-th and $\mR{}$-th states.
By solving $E_{{\rm L},\mL{}}(X)=E_{{\rm R},\mR{}}(X)$, 
we find the degeneracy point
\begin{equation}
  \label{eq:Xcross}
  X_{\mL{},\mR{}} \equiv \frac{\mL{}}{\mL{}+\mR{}} L
  .
\end{equation}
Since $E_{{\rm L},\mL{}}(X)$ and $E_{{\rm R},\mR{}}(X)$ monotonically depend  
on $X$, 
$E_{{\rm R},\mR{}}(X) \gtrless E_{{\rm L},\mL{}}(X)$ 
holds 
if $X \gtrless X_{\mL{},\mR{}}$.

In the following, we assume that 
there is no other level crossing that involves 
the eigenstates $\mL{}$ and $\mR{}$ in the second process 
$x_0 \le X \le x_1$.
This condition is 
\begin{gather}
  \max(X_{\mL{}-1,\mR{}}, X_{\mL{},\mR{}+1})
  < x_0 < 
  X_{\mL{},\mR{}}
  \nonumber\\ 
  < x_1 <
  \min(X_{\mL{}+1,\mR{}}, X_{\mL{},\mR{}-1})
  ,
\end{gather}
and $x_0 < X_{\mL{},\mR{}} < x_1$.

Now we determine the quantum number $n$ 
for $\mL{}$-th and $\mR{}$-th states at the initial point of 
the first process, where $X=x_0$ holds.
Let $\nL$ and $\nR$ denote the values of $n$ for
$\mL{}$-th and $\mR{}$-th states, respectively.
From the condition for $x_0$ examined above, 
we find $\nL=\nR+1$.
In the left (right) well, there are $\mL{}-1$ ($\mR{}-1$) stationary states
below the $\mL{}$($\mR{}$)-th stationary state.
Hence, we obtain $\nR=(\mL{}-1)+(\mR{}-1)+1 = \mL{}+\mR{}-1$, and
$\nL=\mL{}+\mR{}$.

On the other hand, at the final point of the process (3), i.e., at $X=x_1$,
the order of the $\mL{}$-th and $\mR{}$-th states is reversed,
i.e., $\nL+1=\nR$, which implies
$\nL=\mL{}+\mR{}-1$ and $\nR= \mL{}+\mR{}$.

The conditions for $x_0$ and $x_1$ can be simplified 
%{\AT1%
if
%when
%}%\AT1%
we specify the crossing point $X_{\mL{},\mR{}}$ by
its initial lower quantum number $n$ as 
$\mR{}=\lfloor{(n+1)/2}\rfloor$ and $\mL=n+1-\mR$, for example,
%AT2
where $\lfloor{x}\rfloor$ is the maximum integer less than $x$.
%AT2
When $n$ is odd, we find
\begin{align}
  \frac{n+1}{2(n+2)}L < x_0 < \frac{1}{2}L < x_1 < 
  \frac{n+3}{2(n+2)}L
  .
\end{align}
On the other hand, when $n$ is even, we find
\begin{align}
  \frac{1}{2}L < x_0 <
  \frac{n+2}{2(n+1)}L 
  < x_1 <
  \frac{n+4}{2(n+2)}L 
  .
\end{align}

\begin{figure}[bt]
  \centering
  \includegraphics[%
	width=7.0cm%
	%width=8.7cm% for double columns
        ]{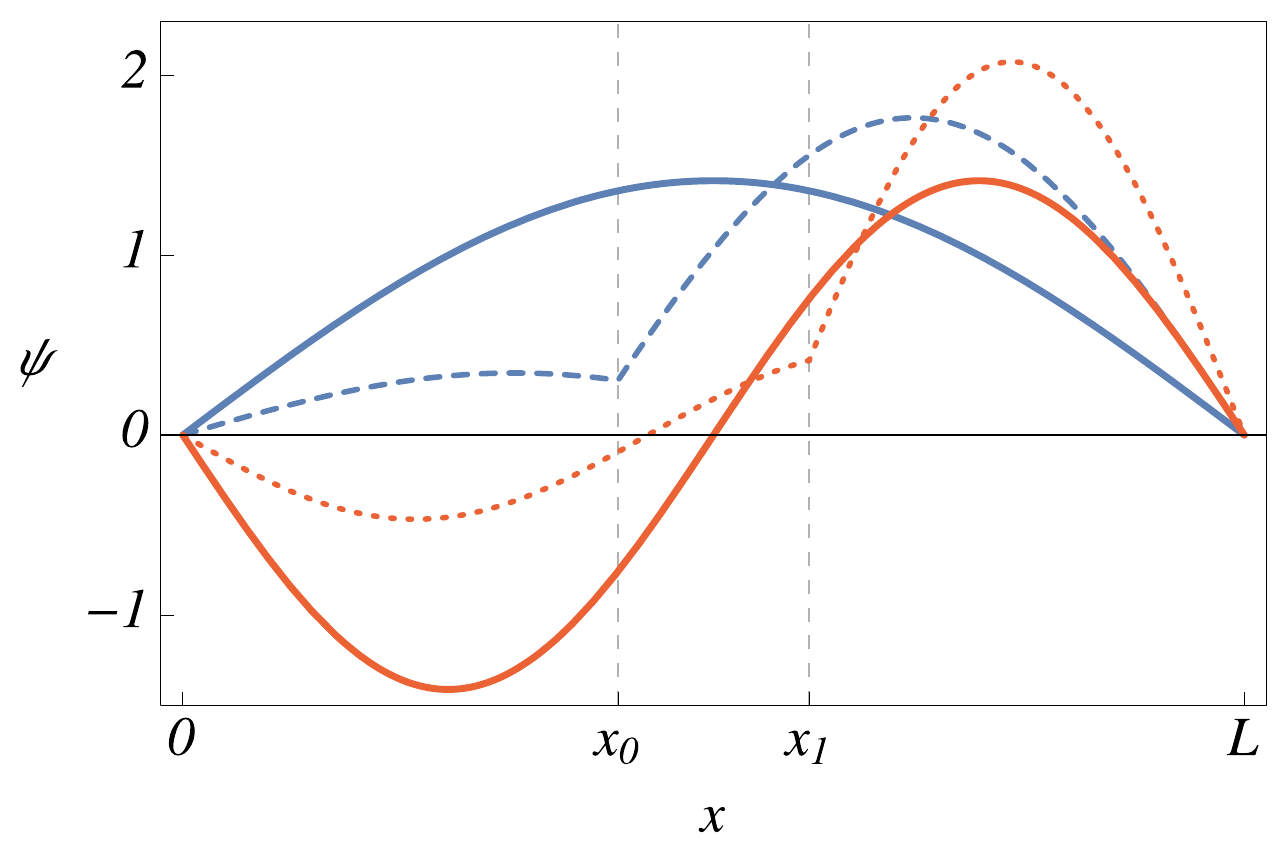}
  \caption{
    (Color online)
    Parametric evolution of 
    %{\AT1%
    an eigenfunction
    %eigenfunctions
    %}%\AT1%
    along $\CX(x_0, x_1)$.
    During the insertion process, the initial eigenfunction $\bracket{x}{1}$ 
    (nodeless, full line) at $g=0$
    becomes $\bracket{x}{1(10g_*, x_0)}$ (dashed), 
    where $g_*=\pi\hbar^2/(2L)$,
    and is finally confined 
    at the right box (not shown here). 
    During the removal process, the initial eigenfunction localized 
    in the right box ($x_1 < x < L$, not shown here) becomes
    $\bracket{x}{2(10 g_*,x_1)}$ (dotted).
    The final eigenfunction is $\bracket{x}{2}$ (with a node, full line).
    The positions of the $\delta$-wall at the insertion and removal are
    indicated by dashed vertical lines.
    We choose $L=1$.
    Other parameters are the same as in Fig.~\ref{fig:n=1}.
  }
  \label{fig:eigenfunction}
\end{figure}

We summarize the argument above to describe the adiabatic evolution 
along the cycle $\CX(x_0,x_1)$. 
The system is prepared to be $\ket{n}$ initially.
We utilize the crossing point $X_{\mL{},\mR{}}$ at the second process, 
where $1\le \mR{}\le n$ and $\mL{}= n+1-\mR{}$ is assumed.
During the first process, the system is in $\ket{n(g,x_0)}$ 
up to the phase factor.
Hence the system is in $n$-th excited state.
At the end of the first process, 
the state $\ket{n(g,x_0)}$ is $\mR{}$-th state in the right well.
Also, during the second process, the state 
keeps to stay 
in $\mR{}$-th state in the right well.
At the same time, the systems is in $(n+1)$-th state of the whole system.
During the third process, 
the system is in $\ket{n+1(g,x_1)}$ up to the phase factor.
Hence, the final state of the cycle is $\ket{n+1}$.
In this sense, the adiabatic excitation from $\ket{n}$ to $\ket{n+1}$
is completed. We depict the examples that adiabatically 
connects the ground and 
first excited states in Figs.~\ref{fig:n=1} and~\ref{fig:eigenfunction}.

We note that, from the construction, the repetition of 
the cycle $\CX(x_0,x_1)$ two times, $\ket{n}$ is delivered to
the initial state $\ket{n}$,
so is $\ket{n+1}$.
Hence the inverse of $\CX(x_0,x_1)$ delivers $\ket{n+1}$ to $\ket{n}$.

\begin{figure}[bt]
  \centering
  \includegraphics[%
        width=7.0cm%
	%width=8.7cm% for double columns
        ]{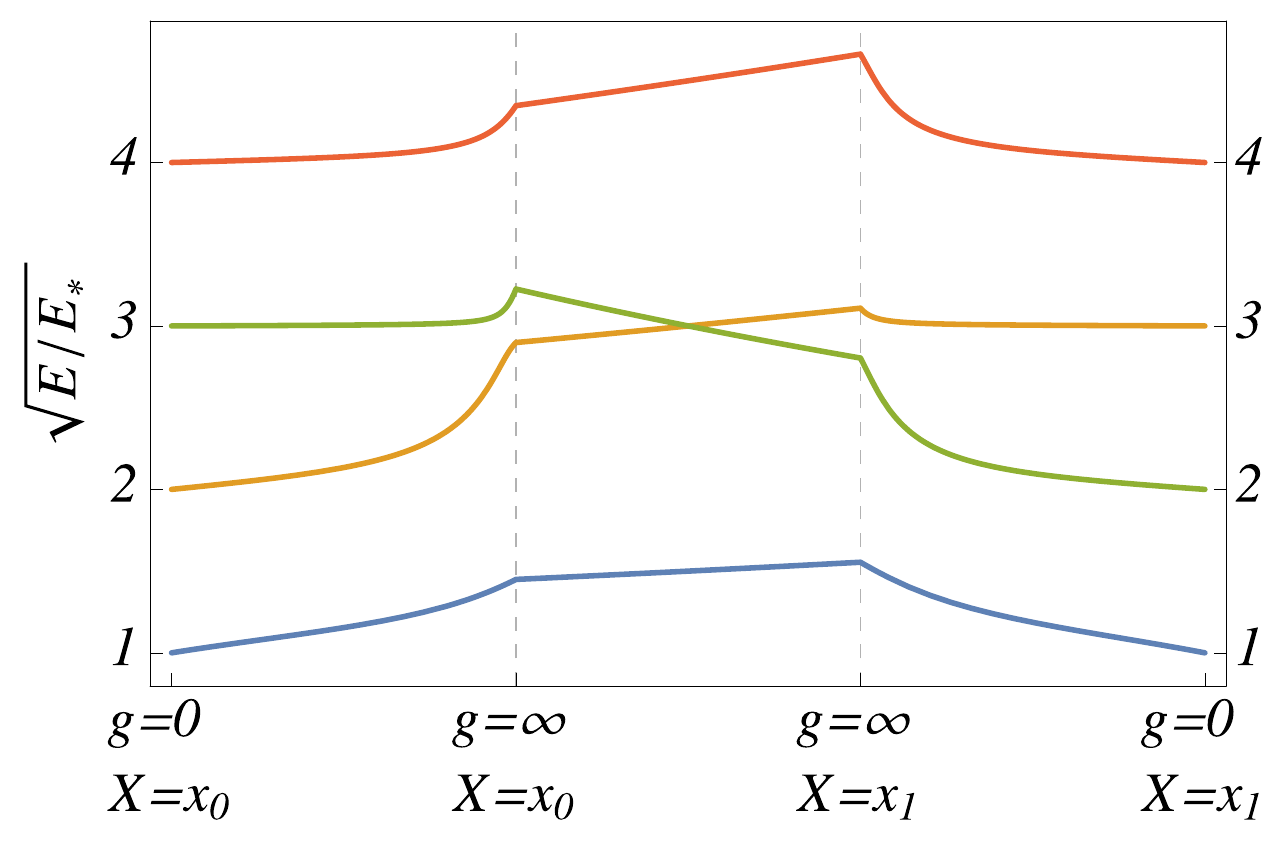}
  \caption{
    (Color online)
    Parametric evolution of eigenenergies along $\CX(x_0, x_1)$
    with $x_0 = 0.31 L$ and $x_1= 0.36 L$. 
    While the eigenenergies of the first and second excited states are
    interchanged, there is no effect on the ground and third excited states.
    }
  \label{fig:other}
\end{figure}

We also note that an arbitrary pair of the eigenstates of $H_0$ can be 
adiabatically connected, if we appropriately combine adiabatic 
cycles $\CX(x_0, x_1)$
%{\AT1%
with various values of $x_0$ and $x_1$.
%}%\AT1
For example, the cycle shown in Fig.~\ref{fig:n=1} connects 
two pairs $(n=1,n=2)$ and $(n=3,n=4)$, whereas the cycle in 
Fig.~\ref{fig:other} connects the pair $(n=2,n=3)$. 
Hence, the ground state $\ket{1}$ can be connected to the 
stationary state $\ket{4}$
by a combination of these two cycles. 
In this sense, these cycles
can be regarded as basis of the adiabatic excitations.

\section{Cycle with $\delta$-wall flip}
\label{sec:CY}
\begin{figure}[tb]
  \centering
  \includegraphics[%
        width=7.0cm%
	%width=8.7cm% for double columns
        ]{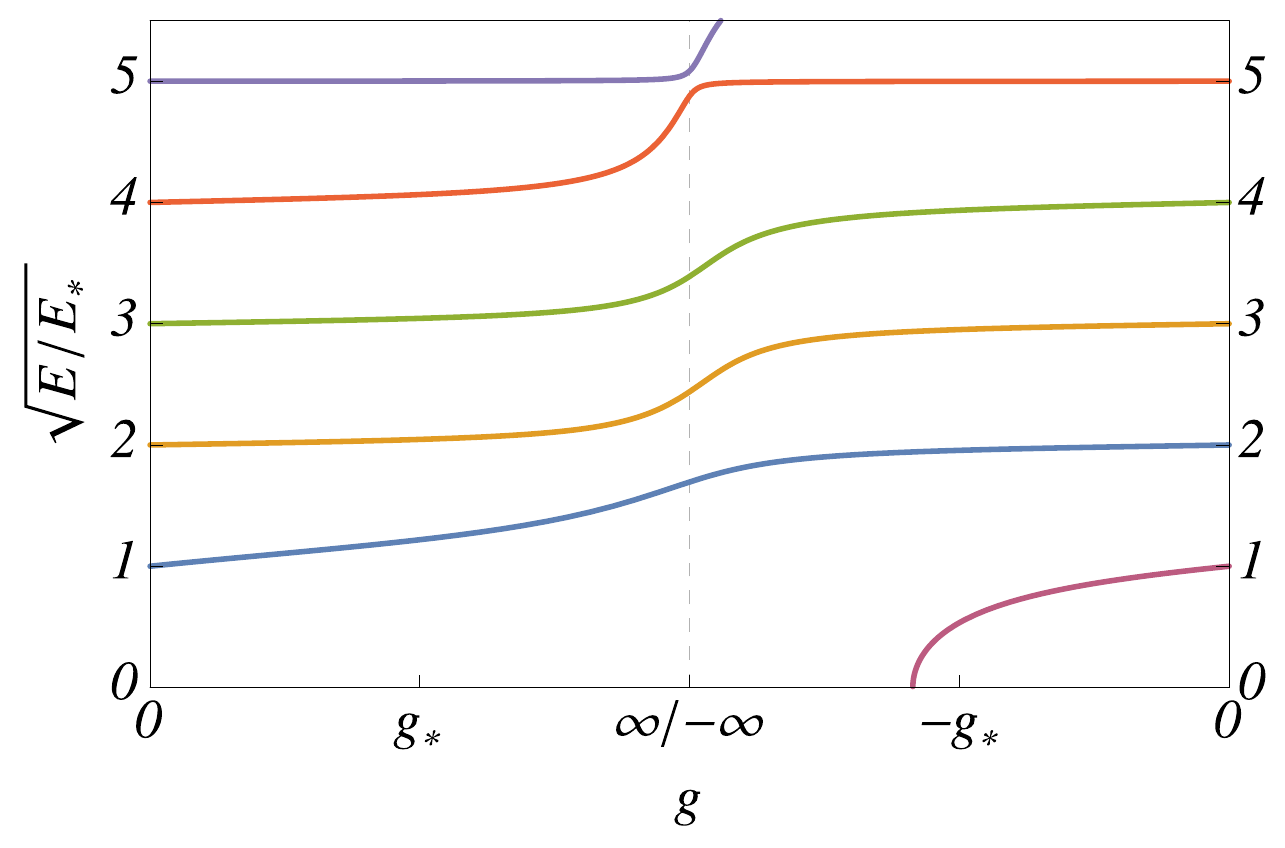}
  \caption{
    (Color online)
    Parametric evolution of eigenenergies along $\CY(x_0)$
    with $x_0 = 0.41 L$.
    The horizontal axis 
    %{\AT1%
      is linear in $\tan^{-1} g$,
      %indicates $g/g_*$, 
    %}%\AT1%
    and we set
    $g_*=\pi\hbar^2/(2L)$.
    The left half and the right half 
    corresponds the first and the third process of $\CY(x_0)$
    respectively. 
    After the completion of $\CY(x_0)$, all eigenenergies are delivered
    to higher neighboring eigenenergies.
    }
  \label{fig:CY}
\end{figure}

We proceed to examine the cycle $\CY(x_0)$ that involves 
an insertion, a flip, and a removal of the $\delta$-wall
placed at $x_0$. 
We will show that $\CY(x_0)$ delivers
an arbitrary stationary state $\ket{n}$ to higher neighboring
state $\ket{n+1}$, if $x_0$ is chosen appropriately. Hence
the resultant permutation of eigenspaces is different from
the one induced by $C_X(x_0, x_1)$.

Our definition of $\CY(x_0)$ is the following. 
First, the strength of the $\delta$-wall $g$ is adiabatically 
increased from $0$ to $\infty$, i.e.,
the first process is the same with $C_X(x_0, x_1)$. 
Second, $g$ is suddenly changed from $\infty$ to $-\infty$ to
flip the wall.
Third, $g$ is adiabatically increased from $-\infty$ to $0$
to remove the wall.

We note that $\CY(x_0)$ resembles a cycle that passes
Tonks-Girardeau and super-Tonks-Girardeau regimes
of the Lieb-Liniger model~\cite{Haller-Science-325-1224}.
In this cycle, the strength of the two-body contact
interaction of Bose particles is varied from zero
to $\infty$, then is changed from $\infty$ to $-\infty$ suddenly,
and is finally increased from $-\infty$ to zero. 
%{\AT1%
The adiabatic cycle excites the system consists of
the Bose particles~\cite{Yonezawa-PRA-87-062113}.
%}%\AT1%
% {\AT1\sout{%
% The nonequilibrium nature of the Bose gas during this cycle
% is experimentally investigated~\cite{Haller-Science-325-1224}.
% The nature of this cycle in a few-body cases is also examined
% in Ref.~\cite{Yonezawa-PRA-87-062113}.
% }}%\AT1\sout{%

We examine the parametric evolution of $E_n(g,X)$ along $\CY(x_0)$
(see, Fig.~\ref{fig:CY}).
As is done in the previous section, we assume that the confinement 
potential is the infinite square well and $x_0/L$ is irrational.

The parametric evolution along the first process is examined in the
previous section: $E_n(g,X)$ monotonically increases with $g$, and 
has no crossing with other levels.

To examine the second process, i.e. the $\delta$-wall flip, 
we utilize the fact that an eigenenergy $E$ satisfies 
a transcendental equation, which is determined by
the connection problem of the eigenfunction at $x_0$~\cite{QMtrans}.
We may examine the transcendental equation with a small parameter
$g^{-1}$~\cite{Ushveridze-JPA-21-955},
%{\AT1%
%}%\AT1%
which concludes that the $n$-th eigenenergy
is connected to the $n^*$-th eigenenergy at $g=-\infty$,
i.e, 
\begin{equation}
  \label{eq:defnast}
  E_n(\infty,x_0)=E_{n^*}(-\infty,x_0),
\end{equation}
where $n^*$ is
an integer.

%{\AT1%
The following proof of $n^*=n+1$ is divided into two parts.
First, we show $n < n^*$. Since $E_n(g,x_0)$ monotonically increases
with $g$ (see, Sec.~\ref{sec:apply}), we obtain
$E^{(0)}_n<E_n(\infty,x_0)$ and $E_{n^*}(-\infty,x_0) < E^{(0)}_{n^*}$.
We find, from Eq.~\eqref{eq:defnast},
$E^{(0)}_n < E^{(0)}_{n^*}$, which implies $n<n^*$.
Second, we show $n^* \le n+1$ by contradiction.
Assuming $n^* > n+1$, we find
$E^{(0)}_{n+1} < E_{n+1}(\infty,x_0) \le E_{n^*}(-\infty,x_0)$ holds
from Eq.~\eqref{eq:Erange}.
Using Eq.~\eqref{eq:defnast}, we obtain $E^{(0)}_{n+1} < E_{n}(\infty,x_0)$,
which contradicts with Eq.~\eqref{eq:Erange}.
%
%Since $E^{(0)}_n < E^{(0)}_{n+1}$ holds,
%we obtain 
%$E^{(0)}_n < E^{(0)}_{n+1} < E_{n}(\infty,x_0),$
%which contradicts with Eq.~\eqref{eq:E_LongRange}.
Namely $n^* \le n+1$ holds.
Thus we conclude $n^*=n+1$.
%}%\AT1
% {\AT1\sout{%
% We show $n^*=n+1$ by contradiction.
% First, suppose $n^* < n+1$ holds. By extending $E_{n^*}(g,x_0)$
% from $g=-\infty$ using Eq.~\eqref{eq:Erange}, 
% we find $E_n^{0} < E_{n^*}^{(0)}$, which contradicts with the
% assumption $n^*< n+1$.
% Second, suppose $n^*> n+1$ holds. This implies that
% either of $E_{n}^{0} \le E_{n+1}^{0} \le E_{n}(\infty,x_0)$, or,
% $E_{n^*}(-\infty,x_0) \le E_{n+1}^{0} \le E_{n^*}^{(0)}$ holds.
% Each case contradicts with the condition~\eqref{eq:E_LongRange}.
% Hence we conclude $n^*=n+1$.
% }}%\AT1\sout{%

The analysis of the third process can be carried out as in the case
for the first process. Hence, we conclude that the eigenenergy 
monotonically increases from $E_{n+1}(-\infty,x_0)$ to $E_{n+1}^{(0)}$
during the third process.

In summary, the adiabatic time evolution along $\CY(x_0)$ is the following.
During the first process, the state is $\ket{n(g, x_0)}$ up to a phase
factor. The flip of the $\delta$-wall do not change the state.
During the third process, the state is $\ket{n+1(g, x_0)}$ up to a phase
factor, and finally arrives at the $(n+1)$-th stationary state
of the unperturbed system.

We remark on the inverse of $\CY(x_0)$. In general,
$\CY^{-1}(x_0)$ adiabatically delivers an arbitrary
stationary state, except the ground state, to the neighboring lower 
stationary state.
On the other hand, if the system is initially in 
the ground state, the corresponding eigenenergy diverges to $-\infty$
as a result of the inverse cycle $\CY^{-1}(x_0)$. The corresponding final 
state is strongly attracted to the $\delta$-wall 
%{\AT1%
  with $g=-\infty$.
%}%\AT1%

%\section{Conclusion}
%{\AT2
\section{Summary and discussion}
%}
\label{sec:conclusion}
We have shown that a confined one-dimensional 
particle are excited by the adiabatic cycles $\CX(x_0,x_1)$ and 
$\CY(x_0)$, where the strength
and the position of a $\delta$-wall is varied. 
Hence we have obtained another simple example of exotic quantum 
holonomy~\cite{Tanaka-PLA-379-1693}.

We have shown a detailed analysis of the case that $V_0(x)$ is
the infinite square well. 
In particular, an appropriate combination of $\CX(x_0,x_1)$ adiabatically 
connects an arbitrary pairs of the stationary states of 
the unperturbed Hamiltonian $H_0$.
%{\AT1%
In this sense, the adiabatic cycles $\CX(x_0,x_1)$ and $\CY(x_0)$ 
can be regarded as basis of the permutations of eigenstates.
As an extension of the present work, 
it may be interesting to find a combination of the cycles
to realize an arbitrary permutation of 
eigenspaces~\cite{Leghtas-JPB-44-154017}.
%}%\AT1%

%{\AT1%
At the same time, we 
%We, however, 
%}%\AT1%
%{\AT2
have shown the basis to extend the present result to the cases with an 
arbitrary confinement potential $V_0(x)$.
In particular, 
an exact analysis of the adiabatic application of the $\delta$-wall is shown,
where the condition for the absence of the level crossing 
during the insert/removal of the $\delta$-wall under an arbitrary
confinement potential $V_0(x)$
is clarified.
Hence,
%}%\AT2
%AT2%emphasize that 
it is straightforward to show that 
the adiabatic excitation is possible for an arbitrary 
confinement potential, as long as 
the unperturbed Hamiltonian has multiple bound states.
The changes required to the present argument depends on 
the position of the nodes of the eigenfunctions of $H_0$.

The present scheme should be experimentally realized within 
the current state of art, e.g., an optical box trap made of
a one-dimensional confinement and two Gaussian 
walls~\cite{Meyrath-PRA-71-041604}.
If an additional Gaussian wall approximates a $\delta$-wall well,
the adiabatic excitation by cycles can be realized.

%{\AT2
Finally, we briefly explain a possible application of the present work 
to produce dark solitons with multiple nodes in a cold atom
Bose-Einstein condensate (BEC).
This is based on the correspondence exploited in Refs.~\cite{Karkuszewski-PRA-63-061601,Damski-PRA-65-013604},
between higher excited states of a single particle system and dark solitons of the BEC.
More precisely, it is shown that a diabatic process, i.e. 
an adiabatic process with a diabatic jump through very narrow level
crossing,
delivers the ground state of a single particle system to its first
excited state, and that we may produce a dark soliton with a single node
using a straightforward extension of the diabatic process to 
the BEC. 
Because of the resemblance with diabatic process studied in Refs.~\cite{Karkuszewski-PRA-63-061601,Damski-PRA-65-013604},
the adiabatic cycles in the present paper may produce 
dark solitons from its many-body ground states, when applied to 
the dilute Bose system.
Moreover, an appropriate combination of the adiabatic cycles will produce
the dark solitons with multiple nodes, which correspond to a higher excited 
state in the single particle system.
% FIXME: some comment on dark solitons with multiple nodes?
%}%\AT2

%%%
%%% Acknowledgments
%%% 
\section*{Acknowledgments}
%\ack
AT wish to thank Professor Taksu Cheon and Professor Kazuo Kitahara 
for discussion.

% iop
%\section*{References}   %NOPREPRINT%

%% for preparation, I prefer to use BibTeX
%\bibliography{local,atsnapshot}

%% for submission, we need to embed the .bbl file here
%\input{local-fixed.bbl} % a workaround
%% Start local-fixed.bbl %%%%%%%%%%%%%%%%%%%%%%%%%%%%%%%%%%%%%%%%%%%%
%merlin.mbs apsrev4-1.bst 2010-07-25 4.21a (PWD, AO, DPC) hacked
%Control: key (0)
%Control: author (72) initials jnrlst
%Control: editor formatted (1) identically to author
%Control: production of article title (-1) disabled
%Control: page (0) single
%Control: year (1) truncated
%Control: production of eprint (0) enabled
%

%% End of local-fixed.bbl %%%%%%%%%%%%%%%%%%%%%%%%%%%%%%%%%%%%%%%%%%%

\end{document}